\documentclass[12pt,headings=big,numbers=noenddot,DIV=14,a4paper]{scrartcl}%

\pdfoutput=1

\usepackage{amsmath,amssymb,amsfonts}
\usepackage[normal,font=small,labelfont=bf,labelsep=period]{caption}
\usepackage[pdftex]{color,graphicx}
\usepackage{subfigure}
\usepackage[english]{babel}
\usepackage[compress]{cite}
\addtokomafont{disposition}{\rmfamily\boldmath}

\usepackage{cancel}

\usepackage[dvipsnames]{xcolor}
\definecolor{darkblue}{rgb}{0,0.2,0.6}
\definecolor{darkgreen}{rgb}{0,0.4,0}

\usepackage[linktoc=page,bookmarks=false,colorlinks=false,
linkbordercolor=RoyalBlue,citebordercolor=ForestGreen,urlbordercolor=CornflowerBlue]{hyperref}

\numberwithin{equation}{section}
\DeclareFontFamily{OT1}{pzc}{}
\DeclareFontShape{OT1}{pzc}{m}{it}{<-> s * [1.10] pzcmi7t}{}
\DeclareMathAlphabet{\mathpzc}{OT1}{pzc}{m}{it}
\title{\bf \LARGE Higgs couplings and electroweak observables:\\ a comparison of precision tests}
\author{\large Riccardo
Barbieri and Andrea Tesi}
\date{\normalsize \it  Scuola Normale Superiore and INFN, Piazza dei Cavalieri 7, 56126 Pisa, Italy}

\begin{document}
\begin{titlepage}

\maketitle
\thispagestyle{empty}

\begin{abstract}
\noindent
Is the weak scale natural? This ever pending question makes the search for new particle production a highly motivated primary goal of the next LHC phase. These searches may or may not be successful. While waiting for a needed higher energy collider to extend the direct exploration, the search for signs of new physics might be confined  to indirect tests for quite some time.  In a few fully calculable models, weakly or semi-strongly interacting, we compare the significance to measure  the Higgs couplings versus the electroweak observables.
\end{abstract}

\vfill
\noindent\line(1,0){188}\\\medskip
\footnotesize{E-mail: \scriptsize\tt{\href{mailto:barbieri@sns.it}{barbieri@sns.it}, \href{mailto:andrea.tesi@sns.it}{andrea.tesi@sns.it}}}
\end{titlepage}

\section{Introduction}
\label{sec1}

A key structural question keeps pending on the foundations of particle physics and of fundamental physics in general: is the weak scale natural?  The results of the first LHC phase, as partly already hinted by previous experiments as well\cite{Barbieri:2000gf}, have shown that the proposed positive answers to this question do not work in the way they were originally thought. To keep the weak scale and the related Higgs boson mass where they are \cite{Aad:2012tfa,Chatrchyan:2012ufa}, some apparently accidental relations between different parameters of the proposed extensions of the Standard Model (SM) have to be invoked, at least at the level of $(1\div 10)\%$. 

At a fundamental level a fine tuning of $(1\div 10)\%$ to make the weak scale totally  insensitive to what happens at shorter distances, whatever they are, even the Planck scale, does not look as a  serious problem, also because fine tunings of this sort are plentiful in nature. Rather the serious problem is of practical nature. While in absence of fine tunings we would have known for sure  where and how the signs of new physics should have been seen - and they have not -, this is no longer the case in the present situation.
Confronted with it, several different attitudes can be and are being taken. Without entering the discussion of pros and contras for any such attitude\footnote{See \cite{Barbieri:2013vca} for a possible list and references.} , here we take the view that it will be in any case crucial for the entire field  to push as high as possible the sensitivity to the signs of ``quasi-natural" theories of ElectroWeak Symmetry Breaking (EWSB), as they may  now be called.

An obvious immediate consequence  is that the search for the production of new particles, as expected in such theories, is a highly motivated primary goal of the next LHC phase. Given the previous recent experience, there are  in fact good reasons to think that these searches will be well in place. The exploration of most part of the sensitive region of parameter space is actually likely to take place in the relatively early stage of the new LHC phase. Another thing can be said quite firmly: the lack of signals so far makes it implausible  that the LHC will be able to explore the full features expected in motivated extensions of the SM, if they are indeed realized.

Here we are concerned with the information that might come from indirect searches of New Physics (NP) in precision measurements. Such measurements could play a leading role in a sufficiently long period of time, after a relatively early stage of the new LHC phase, whatever its findings will be, and before the advent of a needed higher energy hadron collider. Specifically we have in mind the measurements of the Higgs boson couplings at the LHC and the improvements in the ElectroWeak Precision Tests (EWPT) that could be done at  a new Z factory, like at an ILC or at TLEP. A different opportunity  is offered by flavour physics  experiments but it will not concern us here. 

There is no general statement that can be made about the relative importance of Higgs coupling measurements and EWPT, since it depends upon the models (or the category of models) under consideration. On the other hand, we find it useful to try to have a sufficiently broad view of the possible outcomes.
To this end we consider three examples in precisely defined regimes, so as to make possible correspondingly precise statements:
\begin{itemize}
\item[i)] A ``composite" Higgs boson from a semi-strong $SO(5)/SO(4)$ $\sigma$-model \cite{Agashe:2004rs}, linearly realized;
\item[ii)] The Minimal Supersymmetric Standard Model (MSSM) with all s-partners heavy enough that their presence, real or virtual, does not influence the precision observables in a significant way;
\item[iii)] The Next-to-Minimal Supersymmetric Standard Model (NMSSM) with s-partners equally decoupled together with the extra scalar doublet orthogonal to the observed states (the Goldstone or the Higgs bosons)
(see \cite{Ellwanger:2009dp} for a review).
\end{itemize}
Our approach is complementary to already existing studies based on effective lagrangians \cite{Giudice:2007fh,Falkowski:2013dza,Contino:2013kra,Pomarol:2013zra,Dumont:2013wma,Grinstein:2013vsa}, see also \cite{Azatov:2012qz,Gupta:2013zza}.

Also in view of the current bounds, these models provide a significant representation of quasi-natural models of EWSB, even though different specific realizations are possible, that can give rise to different features. The early results of the LHC in its second phase might clearly point to i) or ii)/ iii), perhaps with some needed integration, or could keep them all as open possibilities. 

\section{A ``composite" Higgs boson}
\label{sec2}

The model we consider is defined by the Lagrangian\cite{Barbieri:2007bh}
\begin{equation}
{\cal L}= \frac{1}{2}(D_\mu \Phi)^2- \lambda (\Phi^2-f_0^2)^2 - V(\varphi,\varphi_5),
\end{equation}
where $\Phi$ is a  five-plet of real scalar fields, $D_\mu$ is the covariant derivative with respect to the SM gauge group and $V(\varphi,\varphi_5)$ is a potential that breaks explicitly the $SO(5)$ symmetry of the $\lambda$-dependent term down to $SO(4)$. Under this $SO(4)$  $\Phi= \varphi +  \varphi_5$ where $\varphi$ is quartet, or a complex doublet under $SU(2)_L\times U(1)_Y$, and $\varphi_5$ is a SM singlet. In a non-linearly realized $SO(5)/SO(4)$ $\sigma$-model
the $\lambda$-term is replaced by $\delta(\Phi^2-f^2)$, where $f$ is the decay constant of the (pseudo)-Goldstone boson field $\varphi$. Here we keep a finite coupling $\lambda$ to increase the calculability of the model.

With a specific choice of the potential $V$, e.g.\cite{Contino:2011np}
\begin{equation}\label{pot5}
V(\varphi,\varphi_5)=\alpha f_0^2 \varphi^2 -\beta \varphi^2\varphi_5^2,
\end{equation}
one can compute the vacuum expectation values of $\varphi$ and $\varphi_5$
\begin{eqnarray}
\langle\varphi\rangle^2 &=& \frac{2 f_0^2 (\alpha -\beta ) \lambda }{\beta  (\beta -4 \lambda )}=v^2= (246~\mathrm{GeV})^2,\\
\langle\varphi_5\rangle^2 &=& \frac{f_0^2 (\alpha  (\beta -2 \lambda )-2 \beta  \lambda )}{\beta  (\beta -4 \lambda )},
\end{eqnarray}
as well as the mass and composition of the two physical scalars in $\Phi$. Let us define 
\begin{equation}%SourceDoc 
 \langle\varphi\rangle^2+\langle\varphi_5\rangle^2= f_0^2 \frac{4\lambda-\alpha}{4\lambda-\beta}\equiv f^2,
\end{equation}
so that, when $\lambda\to \infty$, then $f_0\to f$ to recover the non-linear $\sigma$-model description. Let us also define the  mass eigenstates ($h,\sigma$) by 
\begin{equation}
h = \cos{\theta} \varphi +  \sin{\theta}\varphi_5,\quad  \sigma = - \sin{\theta}  \varphi  + \cos{\theta}\varphi_5,
\end{equation}
where we maintain the same notation  $\varphi$ for its only physical component. If one insists that the parameters of the breaking potential, $\alpha, \beta$, remain limited as $\lambda$ grows, the parameters $\alpha, \beta, \lambda$ and $f_0$ can be traded for the more physical parameters $v, f$ and the masses $m_h$, $m_\sigma$ in a unique way, e.g.
\begin{equation}
\lambda= \frac{m_\sigma^2 + m_h^2}{8f^2}.
\end{equation}
 In this way the mixing angle is also uniquely determined by
 \begin{equation}
\sin 2\theta = - 2\sqrt{\xi(1-\xi)}\frac{m_\sigma^2+m_h^2}{m_\sigma^2-m_h^2}\sqrt{1-\frac{m_h^2m_\sigma^2}{(m_\sigma^2+m_h^2)^2 (1-\xi)\xi}},
\label{s2teta}
\end{equation} 
 where we define as customary
 \begin{equation}
\xi =\frac{v^2}{f^2}.
\end{equation} 
For large $m_\sigma^2/m_h^2$ we have
\begin{equation}
 \sin^2\theta =  \xi -\frac{m_h^2}{m_\sigma^2} + O(\xi \frac{m_h^2}{m_\sigma^2}).
\end{equation}
Had we considered a  different $SO(5)$-breaking potential than (\ref{pot5}), e.g. $V=\alpha f_0^3 \varphi_5 -\beta f_0^2\varphi^2$\cite{Barbieri:2007bh},  we would have obtained a similar  expression except for a factor of 2 in front of the $m_h^2/m_\sigma^2$ correction.\footnote{The potential (\ref{pot5}) can be viewed as the linearized version of the Minimal Composite Higgs Model MCHM$_{5,10}$ \cite{Contino:2006qr} with SM fermions coupled linearly to composite fields in the fundamental or antisymmetric representation of $SO(5)$, whereas $V=\alpha f_0^3 \varphi_5 -\beta f_0^2\varphi^2$ represents the linearized version of MCHM$_4$ \cite{Agashe:2004rs} with composite fermions in the spinorial representation.}

In a true strongly interacting scenario, one expects the presence of many other resonances. Here we focus just on the $\sigma$ particle, because, despite its simplicity, this model provides the leading contributions to the observables we are interested in, as we are now going to discuss.

The mixing angle (\ref{s2teta}) is the main parameter that determines both the modified Higgs couplings to the gauge bosons, $V=W,Z$, as well as the corrections to the $\varepsilon$-parameters of the EWPT \cite{Altarelli:1990zd}.  For the Higgs couplings, normalized to the SM one has,\footnote{One can show that, in the $m_\sigma\to\infty$ limit, scattering amplitudes sensitive to the couplings in eq.\eqref{couplings} (e.g. $VV\to VV, hh$) agree with those of the non-linear $\sigma$-model \cite{Contino:2013gna}.}
\begin{equation}
\frac{g_{hVV}}{g_{hVV}^{\rm SM}}= \cos\theta,\quad\quad \frac{g_{hhVV}}{g_{hhVV}^{\rm SM}}= \cos^2\theta
\label{couplings}
\end{equation}
and, for  those ones of the $\sigma$ field,
\begin{equation}
\frac{g_{\sigma VV}}{g_{hVV}^{\rm SM}}= -\sin\theta,\quad\quad \frac{g_{\sigma \sigma VV}}{g_{hhVV}^{\rm SM}}= \sin^2\theta.
\end{equation}
As a consequence, for the $\varepsilon_i, i=1,2,3$
\begin{equation}
\varepsilon_i = \varepsilon^{\mathrm{SM},\cancel{h}}_i+ \cos^2\theta \bar \varepsilon_i(m_h) +\sin^2\theta \bar\varepsilon_i(m_\sigma),
\end{equation}
where $\varepsilon_i^{\mathrm{SM},\cancel{h}}$ are the total SM contributions but the Higgs exchanges, while $\bar\varepsilon_i$ are the pure Higgs contributions to the $\varepsilon$-parameters in the SM (see Appendix). We do not consider modifications of the Higgs-fermions couplings nor the virtual effect of any extra particle other than the $\sigma$-scalar itself.

In the large $m_\sigma$ limit, for the deviations from the SM values $\Delta\varepsilon_i \equiv \varepsilon_i-\varepsilon_i^{\rm SM}$, one gets
\begin{eqnarray}\label{finite-1}
\Delta\varepsilon_1 &=&- \sin^2{\theta} \frac{3\alpha}{8\pi c_w^2} \bigg[\log \frac{m_\sigma}{m_h} + c_1(m_h) + O(\frac{m_Z^2}{m_{\sigma}^2})\bigg],\\
\label{finite-2}
\Delta\varepsilon_2 &=& \sin^2{\theta}  \frac{\alpha}{4\pi c_w^2}\bigg[ c_2(m_h) + O(\frac{m_Z^2}{m_{\sigma}^2})\bigg],\\
\label{finite-3}
\Delta\varepsilon_3 &=& \sin^2{\theta}  \frac{\alpha}{24\pi s_w^2} \bigg[\log \frac{m_\sigma}{m_h} + c_3(m_h) + O(\frac{m_Z^2}{m_{\sigma}^2})\bigg],
\label{eps123}
\end{eqnarray}
where numerically for $m_h = 125$ GeV
\begin{equation}
c_1 = -0.57, \quad\quad
c_2 = 0.10,\quad\quad
c_3 = 0.62.
\label{fin}
\end{equation}
As noticed in \cite{Orgogozo:2012ct}, to obtain the  values of the finite terms $c_i$, one has to include the correct dependence of the $\varepsilon_i$ on $m_h$. To this end, it is worth to stress that $\varepsilon_i$ do not only depend on the vacuum polarization amplitudes entering the usual parameters $S, T, U$ \cite{Peskin:1991sw}, but also on other form factors that cannot be related to the former (see Appendix).
\begin{figure}[t]
\begin{center}
\includegraphics[width=0.5\textwidth]{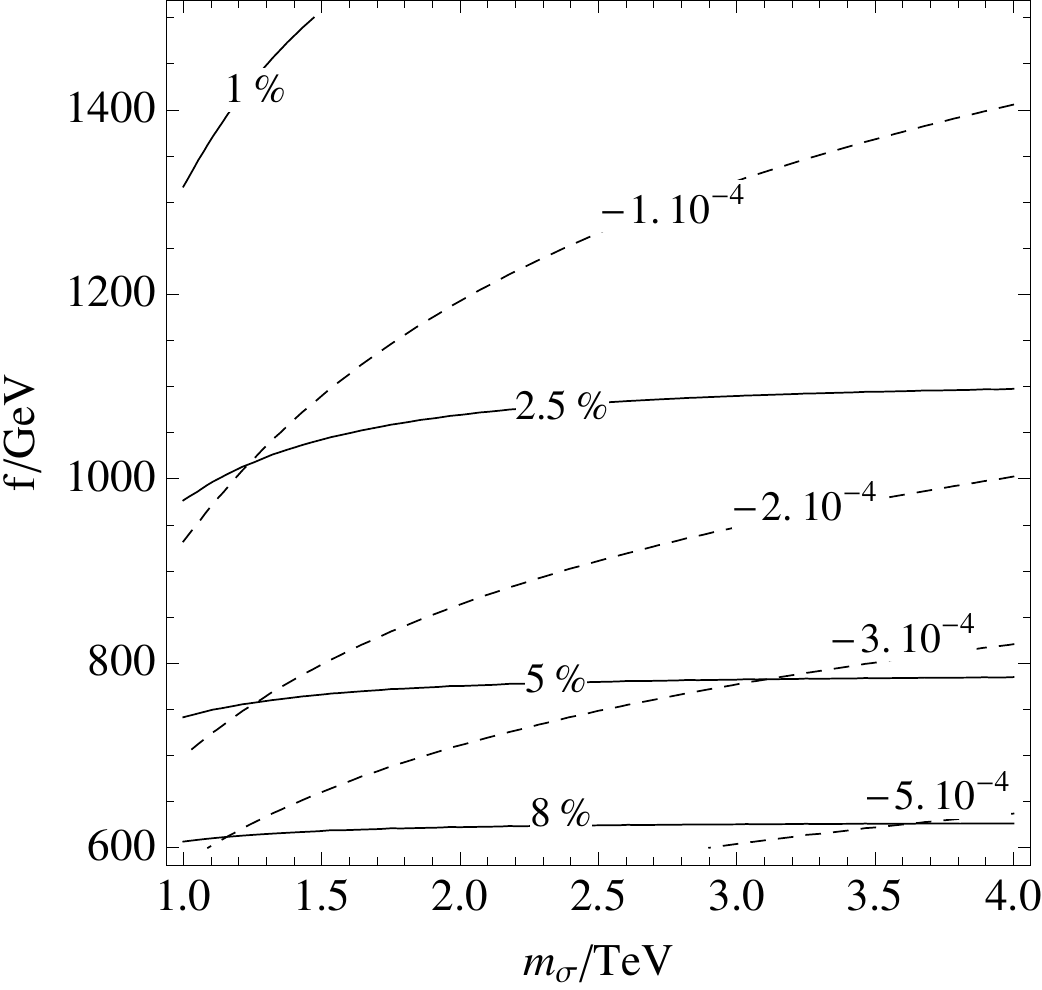}
\caption{\label{fig1}\small ``Composite'' Higgs model. Isoline of $|\delta g_{hVV}|$ (solid) and $\Delta\varepsilon_1$ (dashed).}
\end{center}
\end{figure}

The outcome of these considerations is represented in Fig.~\ref{fig1}, where we show  the relative deviation of $g_{hVV}$ from the SM and the value of $\Delta\varepsilon_1$ as a most representative quantity in the EWPT. In all of the $(m_\sigma, f)$ plane,  $\lambda$ is below 3, i.e. in a semi-perturbative regime, with $\Gamma_\sigma< m_\sigma$. At LHC the $1 \sigma$ attainable precision on
$g_{hVV}$ is expected to be around $5\%$ after $300 ~\mathrm{fb}^{-1}$ and it might be lowered by a factor of about 2 in the High Luminosity configuration (HL-LHC)\cite{ATLAS-HL,CMS:2013xfa} with a corresponding reduction of the theory uncertainties\footnote{See \cite{Almeida:2013jfa}  for a recent detailed analysis.}. A precision below $1\%$ is expected on the other hand in a Higgs factory at an $e^+ e^-$ collider \cite{TLEP-study}. About the EWPT, the error on the parameter $\Delta\varepsilon_1$, currently  of $(5\div 8) 10^{-4}$ depending on the assumptions of the fit \cite{Baak:2012kk,Ciuchini:2013pca}, might be reduced by more than one order of magnitude at TLEP \cite{TLEP-study,Mishima,Ciuchini-et-al}.

\section{The Next-to-Minimal Supersymmetric Standard Model}
\label{sec3}

As a relevant representative of a weakly coupled theory, we consider the NMSSM with s-partners heavy enough that their virtual exchanges do not affect in a significant way the precision observables of interest here.  The reason to consider first the NMSSM than the MSSM is its formal connection with the model discussed in the previous Section: in the limit where the extra scalar doublet orthogonal to the observed states (the Goldstone and the Higgs bosons) is also decoupled, the two residual physical scalars are again an admixture of an $SU(2)$ doublet $H$ and a real singlet $S$.\footnote{The pseudo-scalar component of the complex singlet is decoupled from the system in presence of CP conservation.} This admixture is controlled by the rotation of an angle $\gamma$ that diagonalizes the scalar mass matrix
\begin{equation}
{\cal M}=\left(
\begin{array}{cc}
 m_Z^2 (\frac{1-t_\beta^2}{1+t_\beta^2})^2 + \frac{2 t_\beta^2}{(1+t_\beta^2)^2} \lambda^2 v^2+ \Delta_t^2 & \lambda v M\\
\lambda v M & m_S^2
\end{array}
\right),
\end{equation}
where $\lambda$ is the usual supersymmetric Yukawa coupling of the NMSSM and $\Delta_t$ lumps the main  radiative correction effects that do no decouple in the heavy s-partner limit. The diagonalization of this matrix, trading $M$ and $m_S$ for the physical masses in the order $m_h < m_{h_2}$, gives
\begin{equation}
\sin^2\gamma= \frac{1}{m_{h_2}^2-m_h^2}\bigg[ \frac{2 t_\beta^2}{(1+t_\beta^2)^2}\lambda^2 v^2 + \Delta_t^2 + m_Z^2(\frac{1-t_\beta^2}{1+t_\beta^2})^2 - m_h^2 \bigg].
\label{sin}
\end{equation}
The formal analogy with the previous model makes it such that Eq.s~(\ref{couplings}-\ref{fin}) are also valid here with the replacements $\theta \rightarrow \gamma$ and $\sigma\rightarrow h_2$. The important difference with the composite Higgs model is that in the NMSSM not only the couplings $g_{hVV}$ but also the couplings to all the fermions, $g_{hf\bar{f}}$, are rescaled by a universal factor $\cos{\gamma}$ relative to  the SM ones.

\begin{figure}[t]
\begin{center}
\includegraphics[width=0.5\textwidth]{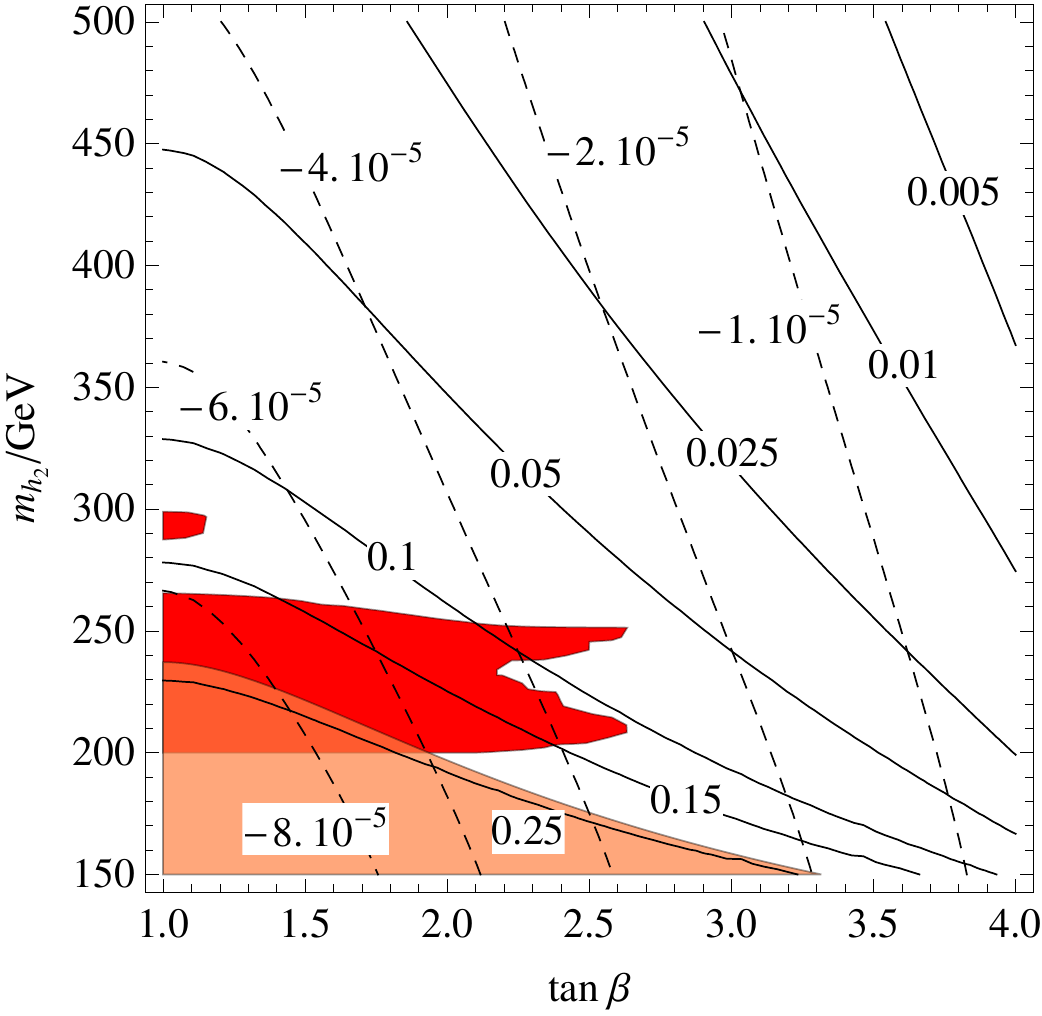}
\caption{\label{fig2}\small NMSSM at $\lambda=0.8$ and $\Delta_t=75$ GeV. Isolines of $\sin^2\gamma$ (solid) and $\Delta\varepsilon_1$ (dashed). The orange region is excluded at 95\%C.L. by the experimental data for the signal strengths of $h$. The red region is excluded by direct searches for $h_2\to ZZ$ \cite{CMS-highmass}. This exclusion above the threshold $h_2\to hh$ depends on the vacuum expectation value of $S$. Here we take $\langle S\rangle= 2 v$.}
\end{center}
\end{figure}

The impact of all this on the precision observables is shown in Fig.~\ref{fig2} for $\lambda = 0.8$, at the upper border for perturbativity up to the Grand Unified Scale \cite{Espinosa:1991gr,Barbieri:2007tu}, and $\Delta_t = 75$ GeV, compatible with stop masses above 700 GeV. How changes in these parameters would affect Fig.~\ref{fig2} is clear from Eq.~(\ref{sin}). In the same figure we also show the currently excluded regions from the measurements of the Higgs couplings and from the direct search of $h_2\rightarrow ZZ$ \cite{CMS-highmass}.

At LHC a universal rescaling by $\cos{\gamma}$ of all the Higgs couplings manifests itself in the signal strengths as an effective  branching ratio in invisible channels. The current limit at $95\%$ C.L., $\sin^2{\gamma} < 0.24$, should be reduced to $\sin^2{\gamma} < 0.15$ after $300 ~\mathrm{fb}^{-1}$ of the next LHC phase, whereas $\sin^2{\gamma} \lesssim 0.05$ might be attainable at HL-LHC \cite{ATLAS-HL,CMS:2013xfa}. An absolute measurement at TLEP of the $hZ$ cross section could increase the sensitivity to $\sin^2{\gamma}$ at the $1\%$ level or less \cite{TLEP-study}. 
Fig.~\ref{fig2} makes clear that the EWPT would have a limited impact on this model.

\section{The Minimal Supersymmetric Standard Model}
\label{sec4}

\begin{figure}[t]
\begin{center}
\includegraphics[width=.5\textwidth]{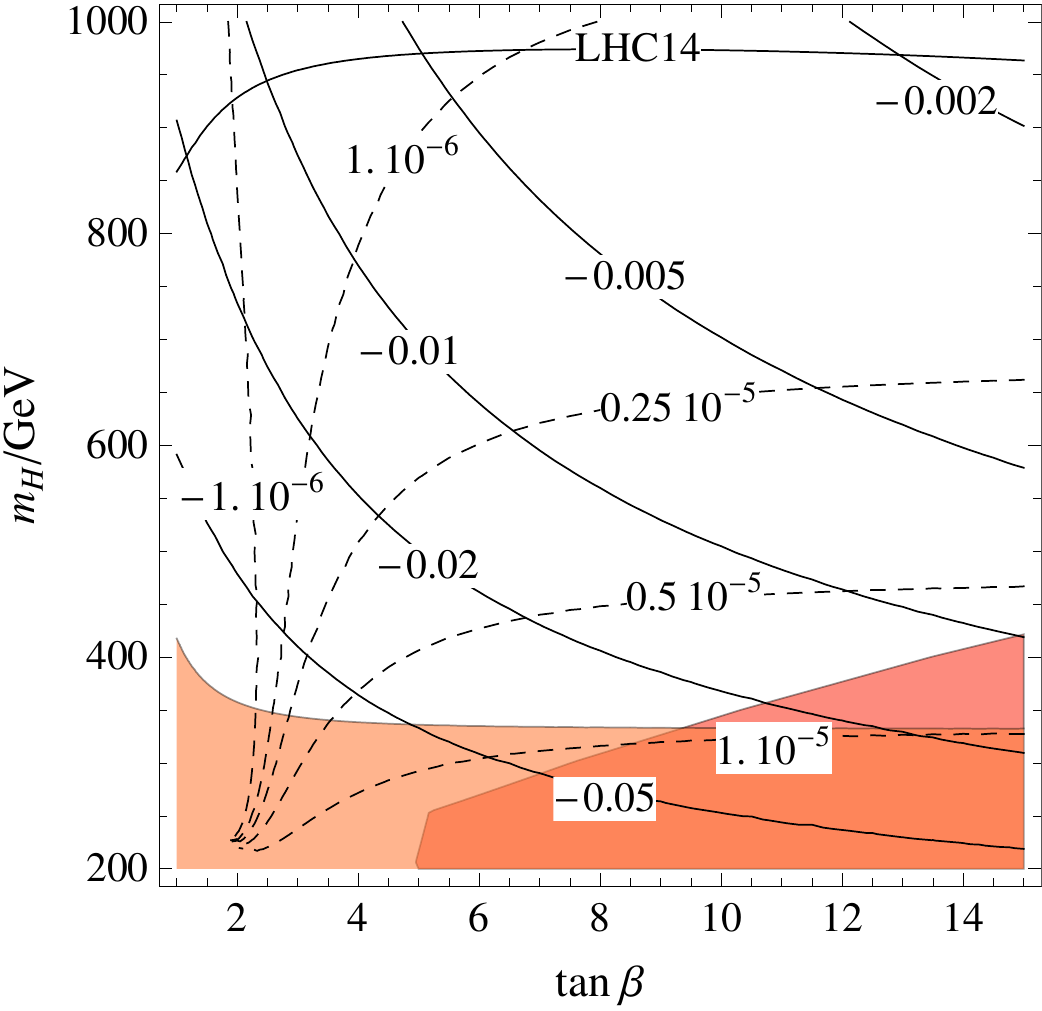}
\caption{\label{fig3}\small MSSM. Isolines of $\sin\delta$ (solid) and $\Delta\epsilon_1$ (dashed). The line LHC14 gives the 95\%C.L. projected exclusion from the sensitivity on the signal strengths of $h$ at ATLAS and CMS with $300 ~\mathrm{fb}^{-1}$.
The orange region is excluded at 95\%C.L. by current data for the signal strengths of $h$. The red region is excluded by CMS direct searches for $A, H\to \tau^+\tau^-$ \cite{CMS:gya}.}
\end{center}
\end{figure}

The MSSM with all s-particles sufficiently decoupled is another relevant example of  a weakly coupled quasi-natural theory of EWSB. The CP-even scalar sector is an admixture of two doublet states: $h_v$, that gets the vacuum expectation value $v$ and its orthogonal combination $h_v^\perp$. For the combination of standard MSSM parameters $(\mu A_t)/\langle m_{\tilde{t}}^2\rangle$ below unity, the mass matrix in the $(h_v, h_v^\perp)$ basis is well approximated by
\begin{equation}\label{massmatrixMSSM}
{\cal M}=\left(
\begin{array}{cc}
m_Z^2 (\frac{1-t_\beta^2}{1+t_\beta^2})^2+ \Delta_t^2 & 2 m_Z^2 \frac{t_\beta\left(1-t_\beta^2\right)}{\left(1+t_\beta^2\right)^2} - \frac{\Delta_t^2}{t_\beta}\\
 2 m_Z^2 \frac{t_\beta\left(1-t_\beta^2\right)}{\left(1+t_\beta^2\right)^2} - \frac{\Delta_t^2}{t_\beta} & m_A^2 +4m_Z^2 \frac{t_\beta^2}{\left(1+t_\beta^2\right)^2} + \frac{\Delta_t^2}{t_\beta^2}
\end{array}
\right).
\end{equation}
 This time one can trade $m_A$ and $\Delta_t$ for the two mass eigenvalues, taken in the order $m_h < m_{H}$, and express in terms of these masses and $t_\beta$ the mixing angle $\delta$, defined by
 \begin{equation}
 h = \cos\delta~ h_v - \sin\delta~ h_v^\perp ,    \quad\quad H = \cos\delta~ h_{v}^\perp + \sin\delta~ h_v.
 \label{mix}
 \end{equation}
 An expression, accurate for $m_{H} \gtrsim 400$ GeV and any value of $t_\beta$, is
  \begin{equation}
\sin\delta = -\frac{m_h^2}{t_\beta m_{H}^2} + \frac{1-t_\beta^2}{1+t_\beta^2} \frac{m_Z^2}{t_\beta m_{H}^2}+O(\frac{1}{m_{H}^4}).
\label{sindelta}
\end{equation}
From Eq.~(\ref{mix}) and the fixed form of the supersymmetric Yukawa couplings, all the Higgs couplings are 
\begin{equation}
\frac{g_{h u\bar{u}}}{g^{\text{SM}}_{h u\bar{u}}}= \cos\delta +\frac{\sin\delta}{\tan\beta}     ,~~\frac{g_{h d\bar{d}}}{g^{\text{SM}}_{h d\bar{d}}}=\cos\delta - \tan\beta \sin\delta,~~\frac{g_{hVV}}{g^{\text{SM}}_{hVV}}= \cos\delta  .
\label{hcouplings}
\end{equation}
\begin{equation}
\frac{g_{Hu\bar{u}}}{g^{\text{SM}}_{hu\bar{u}}}=\sin\delta-\frac{\cos\delta}{\tan\beta},~~\frac{g_{Hd\bar{d}}}{g^{\text{SM}}_{hd\bar{d}}}=\sin\delta+\tan\beta\cos\delta,~~\frac{g_{HVV}}{g^{\text{SM}}_{hVV}}= \sin\delta.
\end{equation}
The isolines of $\sin{\delta}$ in the $(\tan{\beta}, m_{H})$ are shown in Fig.~\ref{fig3}, together with the currently excluded regions, at $95\%$C.L. and within the given assumptions, from the fit of the Higgs couplings and from the search for $A, H\rightarrow \tau \bar{\tau}$ \cite{CMS:gya}.

To determine the sensitivity to  $\sin{\delta}$ in the next LHC phase after $300 ~\mathrm{fb}^{-1}$ of integrated luminosity we use the projected uncertainties of the measurements of the signal strengths of the Higgs boson by ATLAS \cite{ATLAS-Collaboration:2012jwa} and CMS \cite{CMS-projections} given in Table~\ref{tab1}\cite{Barbieri:2013nka}. The corresponding  $95\%$C.L. exclusion line with SM central values is also shown in Fig.~\ref{fig3}. 

\begin{table}[h]
\begin{center}
\begin{tabular}{ccc}
& ATLAS & CMS \\
\hline
$h \to \gamma \gamma$ & 0.16 & 0.15 \\
$h \to Z Z$ & 0.15 & 0.11 \\
$h \to W W$ & 0.30 & 0.14 \\
$V h \to V b \bar{b}$ & -- & 0.17 \\
$h \to \tau \tau$ & 0.24 & 0.11 \\
$h \to \mu \mu$ & 0.52 & -- \\
\end{tabular}
\caption{\label{tab1}Projected uncertainties of the measurements of the signal strengths of $h$ at the 14 TeV LHC with $300~\mathrm{fb}^{-1}$.}
\end{center}
\end{table}

The EWPT observables receive contributions from the complete Higgs system, determined in terms of $\sin{\delta}$ and the masses of all the physical states $m_h, m_{H}, m_A, m_{H^\pm}$. In the formal limit of large $m_{H}, m_A, m_{H^\pm}$ at fixed $\sin{\delta}$ one would obtain the usual ``infrared" logarithms of the same form as in Eq.s~(\ref{eps123}). However, as seen in Eq.~(\ref{sindelta}), $\sin^2{\delta}$ vanishes as $1/m_H^4$. As a consequence the EWPT observables, at $m_H\gtrsim 400$ GeV, are not dominated by the mixing effect, as in the previous cases, but by the non-degeneracy of the $H, A, H^\pm$ states, which gives effects scaling like $1/m_H^2$. The explicit expressions of the $\Delta\varepsilon_i$ at leading order in $1/m_H^2$ are given in Appendix. Numerically one sees the EWPT do not play any role for this configuration of the MSSM.

\section{Discussion of the results}
\label{sec5}

Although with  differences in the different cases, the main conclusion that we can draw, as emerging from Fig.s~\ref{fig1},\ref{fig2} and \ref{fig3}, is that precision measurements will have something significant to say for relevant configurations of every model that we have examined. This is particularly the case for the measurements of the Higgs couplings which will always be able to explore  a significant portion of the different parameter spaces. On the contrary the role of precision measurements of the EW observables, even  pushed  at a dedicated Z-factory, appears limited to the case of a ``composite" Higgs boson.

Coming to the individual cases, the key feature that makes the ``composite" Higgs model particularly sensitive to precision measurements, both of the Higgs couplings and of the EW observables, as shown in Fig.~\ref{fig1}, is the possible separation between the symmetry breaking scale $f$ and the mass of the ``composite" resonances, represented in the linear model by the $\sigma$-particle. In spite of the crudeness of the model, adopted for its calculability, the relation of the scale $f$ with the strength of the linear Higgs couplings to the vectors is not subject to significant model-dependent corrections\cite{Giudice:2007fh}. More model dependent in a truly strongly interacting Higgs boson are the EW observables. Nevertheless the ``infrared logarithms", which are the main feature in Fig.~\ref{fig1}, will anyhow be there\cite{Barbieri:2007bh}. In turn this makes  at least highly unlikely that an improved measurement of, say, the $\varepsilon_1$ parameter, at the level necessary to see an effect like in Fig.~\ref{fig1}, could end up being consistent with the SM value. 

As in the linear $\sigma$-model also the NMSSM can show a mixing of the Higgs boson with an $SU(2)$-singlet scalar, with two important, although formal, differences. One is that the mixing is controlled by the single heavier scale, i.e. the mass of the extra  scalar. (See Eq.~(\ref{sin})). The other difference is that this same mixing suppresses all the couplings of the Higgs boson to the vectors and to the fermions  in the same way. These differences are  at the origin of the relatively weaker explorative power in Fig.~\ref{fig2}, with respect to Fig.~\ref{fig1}, by the precision measurements. An absolute measurement of the invisible Higgs width would be the key here, as possible at an $e^+ e^-$ collider\cite{TLEP-study}. Another possibility is offered by the measurements of the triple Higgs coupling, with conceivable deviations of relative order unity from the SM\cite{Barbieri:2013hxa}, against a $30\%$ $1\sigma$ accuracy foreseen at HI-LHC.

The third case that we have examined is the MSSM with  s-particles sufficiently heavy that their virtual exchange does not influence the precision measurements and with the extra scalars, although heavier than the observed Higgs state, that could be the lightest new particles around. In this case the key features that makes powerful the measurements of the Higgs couplings are: i) their distortion by the mixing  between $h_v$ and $h_v^\perp$, different for vectors, the top quark or the bottom/$\tau$; ii) the dependence of the mixing angle $\delta$ on $m_H$ and $\tan{\beta}$ given in Eq.~\eqref{sindelta} and shown in Fig.~\ref{fig3}.

Given the configuration of the models that we are considering, the competitor of the precision measurements is  the direct search for extra scalars, be they new Higgs particles or some strongly interacting new states. This is manifest, for example,  in Fig.s~\ref{fig2} and \ref{fig3}, where exclusion regions due to direct heavy Higgs searches are already present. It would be interesting to know as reliably as possible the future sensitivity of the LHC, including the high luminosity phase, in the parameter spaces at least of the MSSM and the NMSSM, i.e. in the planes of Fig.s~\ref{fig2} and \ref{fig3},  where in fact the properties of the extra scalars are precisely defined. 
It appears, however, that  the precision measurements will anyhow play an important complementary role.

\subsubsection*{Acknowledgments}

This work is supported in part by the European Programme ``Unification in the LHC Era",  contract PITN-GA-2009-237920 (UNI\-LHC) and by MIUR under the contract 2010 YJ2NYW-010.

\appendix
\section{Computation of the $\varepsilon$-parameters}
\label{app:A}
In this Appendix we collect some reference formulae for the $\varepsilon$-parameters in models with an extra real singlet or an extra $SU(2)$ doublet.
We follow the convention of \cite{Barbieri-Cargese}. The vacuum polarization amplitudes of gauge bosons are
\begin{equation}
\Pi_{ij}^{\mu\nu}(q^2) = -i\big[ A_{ij}(0) + q^2 F_{ij}(q^2) \big] \eta^{\mu\nu} +(q^\mu q^\nu\mathrm{\small -terms}).
\end{equation}
The $\varepsilon$-parameters are related to vacuum polarization amplitudes as\cite{Barbieri:1991qp}
\begin{eqnarray}\label{epsilon}
\varepsilon_1 &=& e_1 - e_5 + \mathrm{non-oblique},\\
\varepsilon_2 &=& e_2 - s_w^2 e_4 - c_w^2 e_5 + \mathrm{non-oblique},\\
\varepsilon_3 &=& e_3 +c_w^2 e_4 - c_w^2 e_5 + \mathrm{non-oblique},
\end{eqnarray}
where
\begin{equation} \label{ei_definitions}
\begin{split}
e_1 &= \frac{A_{33}(0)-A_{WW}(0)}{m_W^2} \, , \\
e_2 &= F_{WW}(m_W^2)-F_{33}(m_Z^2) \, , \\
e_3 &= \frac{c_w}{s_w}F_{30}(m_Z^2) \, ,
\end{split} \qquad
\begin{split}
e_4 &= F_{\gamma\gamma}(0)-F_{\gamma\gamma}(m_Z^2) \, , \\
e_5 &= m_Z^2F_{ZZ}'(m_Z^2) ,
\end{split} 
\end{equation}
and $s_w=\sin\theta_w$ and $c_w=\cos\theta_w$, with $\theta_w$ the weak mixing angle. 
For later convenience we write each SM $\varepsilon$-parameter (and $e_i$) as sum of two different contributions
\begin{equation}
\varepsilon^{\mathrm{SM}}_i= \varepsilon^{\mathrm{SM},\cancel{h}}_i + \bar \varepsilon_i,
\end{equation}
\begin{equation}
e^{\mathrm{SM}}_i= e^{\mathrm{SM}, \cancel{h}}_i + \bar e_i,
\end{equation}
where  the second term is the purely Higgs contribution  while the first one is the rest. Notice that with this definition some of the $\bar e_i$ are individually divergent. In the following we are interested in the NP contribution, defined as
\begin{equation}
\Delta\varepsilon_i = \varepsilon_i-\varepsilon^{\mathrm{SM}}_i
\end{equation}
as well as in $\Delta e_i$ defined in the same way.

\subsection{Extra real singlet}
Following the notation of Section \ref{sec2}, in this case we have
\begin{equation}
\Delta e_i= \sin^2\theta \big[\bar e_i (m_\sigma) -\bar e_i(m_h)\big].
\end{equation}
Any net contribution to $\Delta e_i$ comes from loop diagrams with the exchange of one vector and one scalar. The computation relies on the SM expression $\bar{e}_i$, explicitly (see \cite{Novikov-et-al})
\begin{eqnarray}\label{ei_asymptotic}
\bar e_1 &=& A_1 \log \frac{m_\sigma}{m_Z} + B_1 + C_1 \frac{m_Z^2}{m_\sigma^2} + O(\frac{m_Z^4}{m_\sigma^4}),\\
\bar e_2 &=& C_2 \frac{m_Z^2}{m_\sigma^2} + O(\frac{m_Z^4}{m_\sigma^4}),\\
\bar e_3 &=&  A_3 \log \frac{m_\sigma}{m_Z} + B_3 + C_3 \frac{m_Z^2}{m_\sigma^2} + O(\frac{m_Z^4}{m_\sigma^4}),\\
\bar e_5 &=& C_5 \frac{m_Z^2}{m_\sigma^2} + O(\frac{m_Z^4}{m_\sigma^4}),
\end{eqnarray}
where the coefficients are
\begin{equation}\label{coeff}
\begin{split}
A_1 &= -\frac{3\alpha}{8\pi c_w^2},\\
A_3 &= \frac{\alpha}{24\pi s_w^2},
\end{split}\quad\quad
\begin{split}
C_1 &= \frac{3 \alpha}{8\pi s_w^2 c_w^2} \big[(1-c_w^4)\log\frac{m_Z}{m_\varphi} -c_w^4 \log c_w \big],\\
C_2 &= \frac{-17 \alpha}{192\pi},\\
C_3 &= -\frac{17 \alpha}{192 \pi s_w^2},\\
C_5 &= \frac{\alpha}{192\pi s_w^2 c_w^2}.
\end{split}
\end{equation}
The $B_i$'s are $m_\sigma$-independent divergent terms that cancel out in physical expressions. The proximity of $m_h$ to $m_Z$ makes it numerically relevant to include the full $m_h$-dependence in the $\bar e_i(m_h)$.
\subsection{Extra complex doublet}
In the notation of Section \ref{sec4}, each $\Delta e_i$ is the sum of two contributions,
\begin{equation}\label{ei-mssm}
\Delta e_i = \sin^2\delta \big[\bar e_i (m_H) -\bar e_i(m_h)\big] +\delta e_i,
\end{equation}
where the first is the usual term due to modified Higgs couplings, whereas the second comes mainly from diagrams with exchange of the $H,A,H^\pm$ scalars, which are sensitive to their splittings. 

Notice that $\delta e_i$ is not vanishing when $\sin\delta = 0$. As an example of this consider that, at tree-level, $m_{H^\pm}^2=m_A^2+m_W^2$ independently of $\sin\delta$. This kind of splitting can be traced back to quartic terms in the scalar potential which feel the EWSB.

Given the expression for $\bar{e}_i$, in the decoupling limit \eqref{sindelta} the first term of \eqref{ei-mssm} is of order $1/m_H^4$. Therefore only $\delta e_i$ gives the leading $O(1/m_H^2)$ contribution to the electro-weak parameters.

From an explicit computation of all the relevant diagrams, we find for the MSSM
\begin{eqnarray}
\label{e1-mssm}
\Delta e_1 &=& \frac{\alpha}{48\pi s_w^2}\frac{m_W^2-\Delta m^2}{m_H^2} +O(\frac{m_W^4}{m_H^4}) ,\\
\label{e2-mssm}
\Delta e_2 &=& -\frac{\alpha}{240\pi c_w^2}\frac{m_W^2}{m_H^2} +O(\frac{m_W^4}{m_H^4}),\\
\label{e3-mssm}
\Delta e_3 &=& \frac{\alpha}{96\pi s_w^2}\frac{\Delta m^2-2 m_W^2}{m_H^2} +O(\frac{m_W^4}{m_H^4}) ,\\
\label{e4-mssm}
\Delta e_4 &=& \frac{\alpha}{120\pi c_w^2}\frac{m_W^2}{m_H^2} +O(\frac{m_W^4}{m_H^4}),\\
\label{e5-mssm}
\Delta e_5 &=& \frac{\alpha(1+t_w^4)}{240\pi s_w^2}\frac{m_W^2}{m_H^2} +O(\frac{m_W^4}{m_H^4}),
\end{eqnarray}
where we have defined $\Delta m^2$ as the $O(1/m_H^2)$ splitting between $m_A^2$ and $m_H^2$
\begin{equation}
\Delta m^2 = \frac{m_h^2}{t_\beta^2} + \frac{m_Z^2(3t_\beta^2-1)}{t_\beta^2(1+t_\beta^2)},
\end{equation}
as one can check diagonalizing \eqref{massmatrixMSSM}.

Two main conclusions stem from the above formulae. First, notice that the leading contribution to $\Delta e_{1,3}$ comes from the $\Delta m^2$ splitting, whereas $\Delta e_2$ is not sensitive to it and vanishing in the custodial limit, \textit{i.e.} $\Delta e_2$ (or $U$) feels $\Delta m^2$ only at $O(1/m_H^4)$ \cite{Haber:2010bw}. Second, the size of $\Delta e_{4,5}$ is comparable with that of $\Delta e_{1,2,3}$, \textit{i.e.} with the Peskin-Takeuchi parameters. Differently from the case of the singlet here also $\Delta e_4$ is relevant in the computation of $\varepsilon_{2,3}$ because of the presence of $H^{\pm}$. 
The asymptotic formulae \eqref{e1-mssm}-\eqref{e5-mssm} are well justified in most of the parameter space of Fig.~\ref{fig3}, where, however, $\Delta \varepsilon_1$ is computed without making the large-$m_H$ approximation.


\begin{thebibliography}{99}  

\bibitem{Barbieri:2000gf}
  R.~Barbieri and A.~Strumia,
  %``The 'LEP paradox',''
  hep-ph/0007265.
 

\bibitem{Aad:2012tfa}
  G.~Aad {\it et al.}  [ATLAS Collaboration],
  %``Observation of a new particle in the search for the Standard Model Higgs boson with the ATLAS detector at the LHC,''
  Phys.\ Lett.\ B {\bf 716} (2012) 1
  [arXiv:1207.7214 [hep-ex]].

\bibitem{Chatrchyan:2012ufa}
  S.~Chatrchyan {\it et al.}  [CMS Collaboration],
  %``Observation of a new boson at a mass of 125 GeV with the CMS experiment at the LHC,''
  Phys.\ Lett.\ B {\bf 716} (2012) 30
  [arXiv:1207.7235 [hep-ex]].
 %\cite{Barbieri:2013vca}
\bibitem{Barbieri:2013vca}
  R.~Barbieri,
  %``ElectroWeak theory after the first LHC phase,''
  arXiv:1309.3473 [hep-ph].
 
\bibitem{Agashe:2004rs}
  K.~Agashe, R.~Contino and A.~Pomarol,
  %``The Minimal composite Higgs model,''
  Nucl.\ Phys.\ B {\bf 719} (2005) 165
  [hep-ph/0412089].
  
  
  \bibitem{Ellwanger:2009dp}
  U.~Ellwanger, C.~Hugonie and A.~M.~Teixeira,
  %``The Next-to-Minimal Supersymmetric Standard Model,''
  Phys.\ Rept.\  {\bf 496} (2010) 1
  [arXiv:0910.1785 [hep-ph]] and references therein.
%%%%%%%%%%%%%%%%%%%%%%%%%%%

\bibitem{Giudice:2007fh}
  G.~F.~Giudice, C.~Grojean, A.~Pomarol and R.~Rattazzi,
  %``The Strongly-Interacting Light Higgs,''
  JHEP {\bf 0706} (2007) 045
  [hep-ph/0703164].

    
\bibitem{Falkowski:2013dza}
  A.~Falkowski, F.~Riva and A.~Urbano,
  %``Higgs At Last,''
  arXiv:1303.1812 [hep-ph].  
 \bibitem{Contino:2013kra}
  R.~Contino, M.~Ghezzi, C.~Grojean, M.~Muhlleitner and M.~Spira,
  %``Effective Lagrangian for a light Higgs-like scalar,''
  JHEP {\bf 1307} (2013) 035
  [arXiv:1303.3876 [hep-ph]].
  
  \bibitem{Pomarol:2013zra}
  A.~Pomarol and F.~Riva,
  %``Towards the Ultimate SM Fit to Close in on Higgs Physics,''
  arXiv:1308.2803 [hep-ph].
\bibitem{Dumont:2013wma}
  B.~Dumont, S.~Fichet and G.~von Gersdorff,
  %``A Bayesian view of the Higgs sector with higher dimensional operators,''
  JHEP {\bf 1307} (2013) 065
  [arXiv:1304.3369 [hep-ph]].
\bibitem{Grinstein:2013vsa}
  B.~Grinstein, C.~W.~Murphy and D.~Pirtskhalava,
  %``Searching for New Physics in the Three-Body Decays of the Higgs-like Particle,''
  JHEP {\bf 1310} (2013) 077
  [arXiv:1305.6938 [hep-ph]].

\bibitem{Azatov:2012qz}
  A.~Azatov and J.~Galloway,
  %``Electroweak Symmetry Breaking and the Higgs Boson: Confronting Theories at Colliders,''
  Int.\ J.\ Mod.\ Phys.\ A {\bf 28} (2013) 1330004
  [arXiv:1212.1380].

%\cite{Gupta:2013zza}
\bibitem{Gupta:2013zza}
  R.~S.~Gupta, H.~Rzehak and J.~D.~Wells,
  %``How well do we need to measure the Higgs boson mass and self-coupling?,''
  Phys.\ Rev.\ D {\bf 88} (2013) 055024
  [arXiv:1305.6397 [hep-ph]].
  %%CITATION = ARXIV:1305.6397;%%
  %11 citations counted in INSPIRE as of 27 Nov 2013

  \bibitem{Barbieri:2007bh}
  R.~Barbieri, B.~Bellazzini, V.~S.~Rychkov and A.~Varagnolo,
  %``The Higgs boson from an extended symmetry,''
  Phys.\ Rev.\ D {\bf 76} (2007) 115008
  [arXiv:0706.0432 [hep-ph]].
%%%%%%%%%%%%%%%%%%%%%%%%%%%%%%%%%%%
%%%%%%%%%%%%%%%%%%%%%%%%%%%%%%%%%%%

\bibitem{Contino:2011np}
  R.~Contino, D.~Marzocca, D.~Pappadopulo and R.~Rattazzi,
  %``On the effect of resonances in composite Higgs phenomenology,''
  JHEP {\bf 1110} (2011) 081
  [arXiv:1109.1570 [hep-ph]].

\bibitem{Contino:2006qr}
  R.~Contino, L.~Da Rold and A.~Pomarol,
  %``Light custodians in natural composite Higgs models,''
  Phys.\ Rev.\ D {\bf 75} (2007) 055014
  [hep-ph/0612048].


\bibitem{Altarelli:1990zd}
  G.~Altarelli and R.~Barbieri,
  %``Vacuum polarization effects of new physics on electroweak processes,''
  Phys.\ Lett.\ B {\bf 253} (1991) 161.
  
  
\bibitem{Contino:2013gna}
  R.~Contino, C.~Grojean, D.~Pappadopulo, R.~Rattazzi and A.~Thamm,
  %``Strong Higgs Interactions at a Linear Collider,''
  arXiv:1309.7038 [hep-ph].
  
    \bibitem{Orgogozo:2012ct}
  A.~Orgogozo and S.~Rychkov,
  %``The S parameter for a Light Composite Higgs: a Dispersion Relation Approach,''
  JHEP {\bf 1306} (2013) 014
  [arXiv:1211.5543 [hep-ph]].


\bibitem{Peskin:1991sw}
  M.~E.~Peskin and T.~Takeuchi,
  %``Estimation of oblique electroweak corrections,''
  Phys.\ Rev.\ D {\bf 46} (1992) 381.

\bibitem{ATLAS-HL}
[ATLAS Collaboration],
ATL-PHYS-PUB-2013-014.

\bibitem{CMS:2013xfa}
  [CMS Collaboration],
  %``Projected Performance of an Upgraded CMS Detector at the LHC and HL-LHC: Contribution to the Snowmass Process,''
  arXiv:1307.7135 [hep-ex].
  
  %\cite{Almeida:2013jfa}
\bibitem{Almeida:2013jfa}
  L.~G.~Almeida, S.~J.~Lee, S.~Pokorski and J.~D.~Wells,
  %``Study of the 125 GeV Standard Model Higgs Boson Partial Widths and Branching Fractions,''
  arXiv:1311.6721 [hep-ph].
  %%CITATION = ARXIV:1311.6721;%%

\bibitem{TLEP-study}
  M.~Bicer, H.~Duran Yildiz, I.~Yildiz, G.~Coignet, M.~Delmastro, T.~Alexopoulos, C.~Grojean and S.~Antusch {\it et al.},
  %``First Look at the Physics Case of TLEP,''
  arXiv:1308.6176 [hep-ex].


\bibitem{Baak:2012kk}
  M.~Baak, M.~Goebel, J.~Haller, A.~Hoecker, D.~Kennedy, R.~Kogler, K.~Moenig and M.~Schott {\it et al.},
  %``The Electroweak Fit of the Standard Model after the Discovery of a New Boson at the LHC,''
  Eur.\ Phys.\ J.\ C {\bf 72} (2012) 2205
  [arXiv:1209.2716 [hep-ph]].
  
\bibitem{Ciuchini:2013pca}
  M.~Ciuchini, E.~Franco, S.~Mishima and L.~Silvestrini,
  %``Electroweak Precision Observables, New Physics and the Nature of a 126 GeV Higgs Boson,''
  JHEP {\bf 1308} (2013) 106
  [arXiv:1306.4644 [hep-ph]].

\bibitem{Mishima}
S. Mishima, talk at the ``Sixth TLEP Workshop'',\\
http://indico.cern.ch/contributionDisplay.py?contribId=30$\&$sessionId=1$\&$confId=257713

\bibitem{Ciuchini-et-al}
 M.~Ciuchini et al., in preparation.

  
\bibitem{Espinosa:1991gr}
  J.~R.~Espinosa and M.~Quiros,
  %``On Higgs boson masses in nonminimal supersymmetric standard models,''
  Phys.\ Lett.\ B {\bf 279} (1992) 92.

%\cite{Barbieri:2007tu}
\bibitem{Barbieri:2007tu}
  R.~Barbieri, L.~J.~Hall, A.~Y.~Papaioannou, D.~Pappadopulo and V.~S.~Rychkov,
  %``An Alternative NMSSM phenomenology with manifest perturbative unification,''
  JHEP {\bf 0803} (2008) 005
  [arXiv:0712.2903 [hep-ph]].
  %%CITATION = ARXIV:0712.2903;%%
  %44 citations counted in INSPIRE as of 24 Nov 2013
  
\bibitem{CMS-highmass}
[CMS Collaboration], https://twiki.cern.ch/twiki/bin/view/CMSPublic/Hig12024TWiki.


\bibitem{CMS:gya}
  [CMS Collaboration],
  %``Search for MSSM Neutral Higgs Bosons Decaying to Tau Pairs in pp Collisions,''
  CMS-PAS-HIG-12-050.

\bibitem{ATLAS-Collaboration:2012jwa}
  [ATLAS Collaboration],
  %``Physics at a High-Luminosity LHC with ATLAS,''
  ATL-PHYS-PUB-2012-001.


\bibitem{CMS-projections}
  [CMS Collaboration],
  %``Physics at a High-Luminosity LHC with ATLAS,''
  CMS-NOTE-2012-006.

 \bibitem{Barbieri:2013nka}
  R.~Barbieri, D.~Buttazzo, K.~Kannike, F.~Sala and A.~Tesi,
  %``One or more Higgs bosons?,''
  Phys.\ Rev.\ D {\bf 88} (2013) 055011
  [arXiv:1307.4937 [hep-ph]].
  
  \bibitem{Barbieri:2013hxa}
  R.~Barbieri, D.~Buttazzo, K.~Kannike, F.~Sala and A.~Tesi,
  %``Exploring the Higgs sector of a most natural NMSSM,''
  Phys.\ Rev.\ D {\bf 87} (2013) 115018
  [arXiv:1304.3670 [hep-ph]].
  %%CITATION = ARXIV:1304.3670;%%
  %18 citations counted in INSPIRE as of 24 Nov 2013

\bibitem{Barbieri-Cargese}
  R.~Barbieri, CERN-TH-6659-92.
  %``Electroweak precision tests: what do we learn?,''
  
  %\cite{Barbieri:1991qp}
\bibitem{Barbieri:1991qp}
  R.~Barbieri, M.~Frigeni and F.~Caravaglios,
  %``Supersymmetry signals in electroweak precision tests at LEP,''
  Phys.\ Lett.\ B {\bf 279} (1992) 169.
  %%CITATION = PHLTA,B279,169;%%
  %110 citations counted in INSPIRE as of 24 Nov 2013
  


\bibitem{Novikov-et-al}
  V.~A.~Novikov, L.~B.~Okun and M.~I.~Vysotsky,
  %``On the Electroweak one loop corrections,''
  Nucl.\ Phys.\ B {\bf 397} (1993) 35.

  V.~Novikov, L.~Okun, A.~N.~Rozanov and M.~Vysotsky,
  %``Leptop,''
  hep-ph/9503308.



\bibitem{Haber:2010bw}
  H.~E.~Haber and D.~O'Neil,
  %``Basis-independent methods for the two-Higgs-doublet model III: The CP-conserving limit, custodial symmetry, and the oblique parameters S, T, U,''
  Phys.\ Rev.\ D {\bf 83} (2011) 055017
  [arXiv:1011.6188 [hep-ph]].


\end{thebibliography}
\end{document}